\documentclass{emulateapj}
\usepackage{lscape}
\shorttitle{\textit{Spitzer-IRS} Spectroscopy of M82}
\shortauthors{Beir\~ao et al.}
\begin{document}
 
\title{Spatially Resolved \textit{Spitzer-IRS} Spectroscopy of the
Central Region of M82}

\email{pbeirao@strw.leidenuniv.nl}

\author{P. Beir\~ao\altaffilmark{1},
        B. R. Brandl\altaffilmark{1}, 
        P. N. Appleton\altaffilmark{2},
        B. Groves\altaffilmark{1},
        L. Armus\altaffilmark{3}, 
        N. M. F\"orster Schreiber\altaffilmark{4},
        J. D. Smith\altaffilmark{5},
        V. Charmandaris\altaffilmark{6} \& 
        J. R. Houck\altaffilmark{7}} 
\altaffiltext{1}{Leiden Observatory, Leiden University, P. O. Box
9513, 2300 RA Leiden, The Netherlands}
\altaffiltext{2}{NASA Herschel Science Center, California Institute of
Technology, Pasadena, CA 91125}
\altaffiltext{3}{Spitzer Science Center, California Institute of
Technology, Pasadena, CA 91125}
\altaffiltext{4}{Max Planck Institut f\"ur Extraterrestrische Physik,
Garching, Germany}
\altaffiltext{5}{Steward Observatory, University of Arizona, Tucson, 
                 AZ 85721}
\altaffiltext{6}{IESL/Foundation for Research and Technology - Hellas,
  GR-71110, Heraklion, Greece and Chercheur Associ\'e, Observatoire de
  Paris, F-75014, Paris, France}
\altaffiltext{7}{Astronomy Department, Cornell University, 219 Space
                 Sciences Building, Ithaca, NY 14853}

\begin{abstract}
We present high spatial resolution ($\sim 35$~parsec)
$5-38\mu$m spectra of the central region
of M82, taken with the Spitzer Infrared Spectrograph.
From these spectra we determined the fluxes and equivalent widths
of key diagnostic features, such as the [NeII]12.8$\mu$m, [NeIII]15.5$\mu$m, and
H$_2$ S(1)17.03$\mu$m lines, and the broad mid-IR
polycyclic aromatic hydrocarbon (PAH) emission features in six
representative regions and analysed the spatial distribution of these
lines and their ratios across the central region.
We find a good correlation of the dust extinction with the CO 1-0
emission. The PAH emission follows closely the ionization
structure along the
galactic disk. The observed variations of the diagnostic PAH
ratios across M82 can be explained by extinction
effects, within systematic uncertainties. The $16-18\mu$m PAH complex is
very prominent, and its equivalent width is enhanced outwards from the galactic
plane. We interpret this as a consequence of the variation of the UV
radiation field. The EWs of the $11.3\mu$m PAH feature and the H$_2$ S(1)
line correlate closely, and we conclude that shocks in
the outflow regions have no measurable influence on the H$_2$
emission.
The [NeIII]/[NeII] ratio is on average low at $\sim$0.18,
and shows little variations across the plane, indicating that the
dominant stellar population is evolved (5 - 6 Myr) and well distributed.
There is a slight increase of the ratio with distance
from the galactic plane of M82 which we attribute  to
a decrease in gas density. Our observations indicate that the
star formation rate has decreased significantly in the last 5
Myr. The quantities of dust and molecular gas in the central area of the galaxy argue
against starvation and for negative feedback processes, observable through the strong extra-planar outflows.

\end{abstract}



\section{Introduction}

M82 (NGC 3034) is an irregular galaxy located at 3.3 Mpc
\citep{freedman88} in the M81 group. It is the closest starburst
galaxy, seen nearly edge-on, with an inclination angle of about $80\deg$.
At infrared wavelengths it is the brightest galaxy on the sky, with a
total infrared luminosity of $3.8\times10^{10}L_{\odot}$
\citep{colbert99}. Most of its luminosity originates from the inner 500 pc hosting 
intense starburst activity presumably triggered by a tidal
interaction with M81 \citep[e.g.][]{yun93}. Evidence for this
interaction comes from the HI streams which connect M81 to all three
(M82, NGC 3077, and NGC 2976) peculiar members of the inner M81 group
\citep{apple81,yun94}. Recent deep optical images also revealed stars
associated with the HI bridge between M81 and M82 \citep{sun05}.

At the distance of M82, $1\arcsec$ corresponds to 15 parsec, which
allows spatially resolved studies of the starburst region. Evidence of
a stellar bar $\sim1$ kpc long is shown by near-infrared studies
\citep[e.g.][]{telesco91,larkin94}, mid-infrared [NeII]12.8$\mu$m and
millimetric CO emission studies \citep{lo87}. According to
\citet{larkin94} and \citet{achter95}, there is a rotating ring of
ionized gas at a radius of $\sim85$ pc, and on the inner side of a ring of molecular gas at
$\sim210$ pc. Two possible spiral arms were also identified by
\citet{shen95} and \citet{mayya05}, at radii of $\sim125$ pc and
$\sim400$ pc. The starburst of M82 drives a bipolar mass outflow out to
several kiloparsecs perpendicular to the plane of the galaxy,
especially evident in X-ray and H$\alpha$
\citep{bregman95,shop98,lehnert99,cappi99,strick04}. Dust has also
been detected in the outflow region
\citep{alton99,heckman00,hoopes05,engel06}. The star forming regions of M82 are
predominantly clustered in the volume enclosed by the molecular gas ring,
indicated by the {\sc HII} region tracers, like the [NeII]12.8$\mu$m
line and the mid- and far-infrared continuum emission
\citep{telesco91,walter02,larkin94,achter95,lipscy04}. Near-infrared hydrogen
recombination lines also arise in these regions, but they are a more
ambiguous tracer of star formation, as they can also be excited by
shocks, though these are unlikely to dominate.

Near-infrared integral field spectroscopy and ISO-SWS mid-infrared
spectroscopy by \citet{schreiber01} allowed a detailed modelling of
starburst activity in the central region \citep{schreiber03}. These
models are consistent with the occurrence of starburst activity in two
successive episodes, about 10 and 5 Myr ago, each lasting a few
million years. However, the spatial studies by \citet{schreiber01} and
others covered only near-infrared wavelengths.  The large
aperture of the ISO-SWS provided a continuous $2.4-45 \mu$m spectrum
but covered the whole central region of M82.  ISOCAM-CVF data
\citep{schreiber02} provided better spatial resolution but the spectra only had a
spatial resolution of $R \sim 40$ and were shortward of $15\mu$m.  

\citet{engel06} published Spitzer-IRS low-resolution spectra
of a 1 arcmin wide strip along the minor axis of M82, intersecting the
disk at the eastern side. The spectra, taken as part of the SINGS
Legacy project, were combined with Spitzer 8 and 24 $\mu$m images, and
show that the emission by polycyclic aromatic hydrocarbons (PAH) and H$_2$ molecules extends far out from the disk (to
6 kpc) in both directions.  \citet{engel06} suggest that the
filamentary aromatic-dominated emission represents dust either
expelled from the galaxy as a result of a powerful nuclear superwind,
or that dust is in the halo being lit up by the starburst,
perhaps coexisting with the extensive warm H$_2$ molecules. 
They suggest that this halo dust is probably a leftover from the interaction with M81. 

In this paper we present mid-IR spectral maps at unsurpassed
sensitivity and spatial resolution of the central $\sim0.5$ kpc$^2$ of
M82, covering the main contributors to the bolometric luminosity of
the galaxy. Our goal is to provide a spatial and spectrally detailed description
of the physical conditions within the central $\sim500$ pc of M82, to help us to give an insight on the evolution of the starburst activity in this region. This involves the study of the distribution of the radiation field, gas density, and the physical properties of PAHs. Of particular interest are the spatial variations of the
fine-structure lines, the excitation of the molecular hydrogen, and
the distribution of the PAH molecules. In section 2 we describe the
observations and data reduction, in section 3 we present the data, and
in section 4 we discuss the scientific results, followed by our
conclusions.

\begin{figure}
\epsscale{1.15}
\plottwo{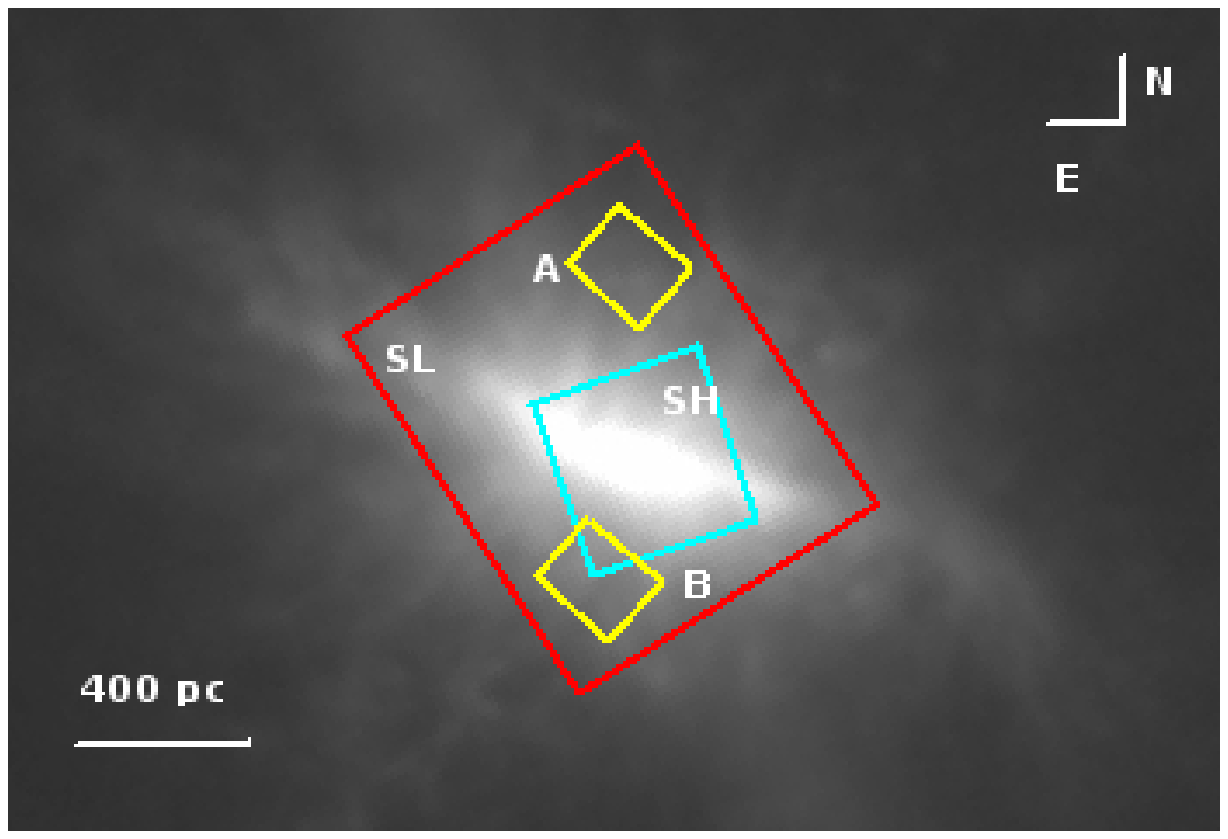}{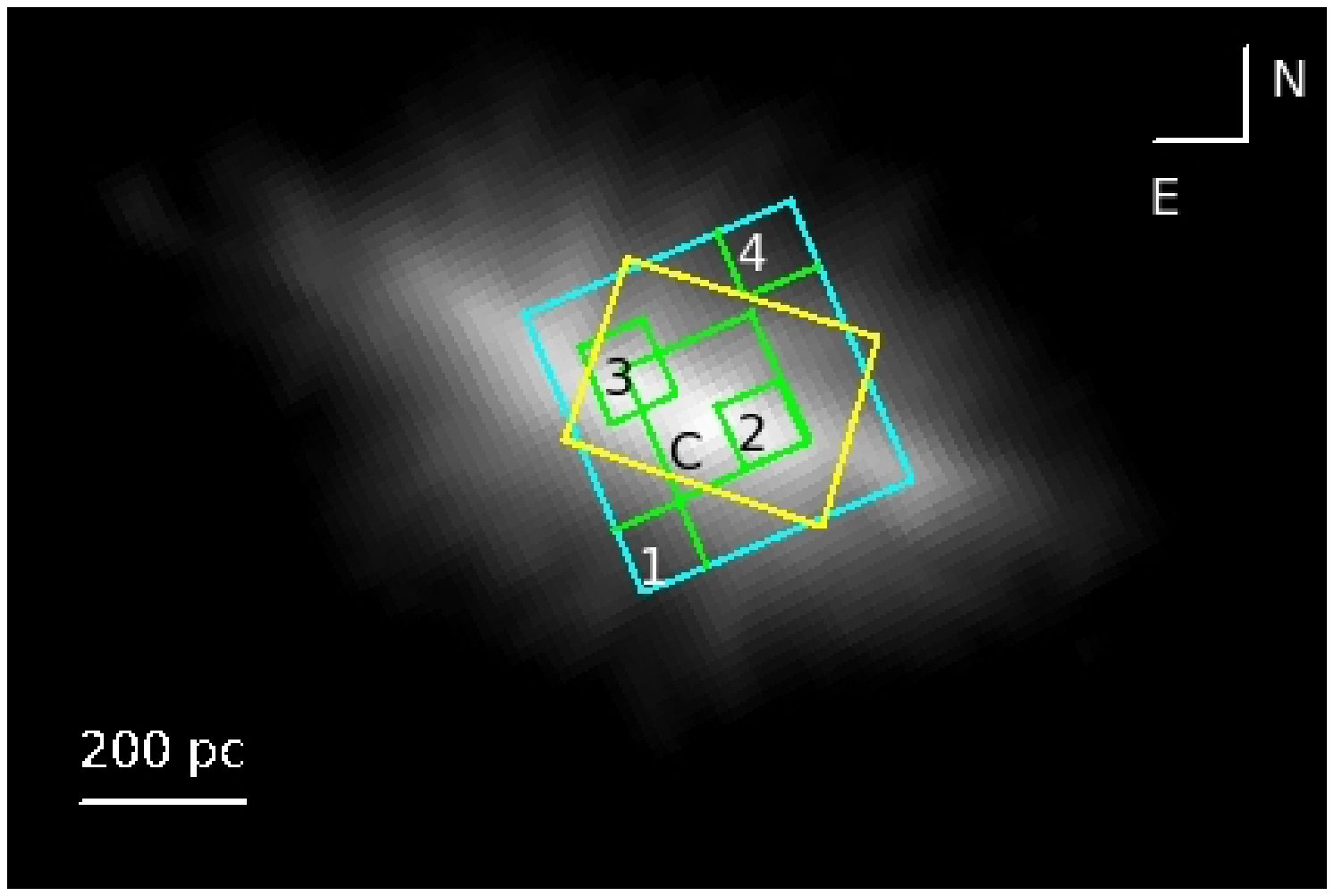}
\caption{Left: Overlay of SL (red) and SH (blue) coverages and selected low-res
         extraction regions on an IRAC 8$\mu$m image. Regions A and B, represented in yellow, are regions where LL (14 - 35 $\mu$m) spectra were extracted. Right: Zoom-in of
         the IRAC 8$\mu$m image with an overlay of the SH map area in blue and the selected regions (in green) from where the SH+SL spectra were extracted. The ISO-SWS aperture used in the $12-27 \mu$m range by
         \citet{schreiber01} is also overlayed in yellow on the image. Both figures are in logarithmic scaling.}
\label{figregions}
\end{figure} 


\section{Observations and Data Reduction}

\begin{deluxetable*}{ccccccccc}
\tabletypesize{\footnotesize}
\tablecolumns{6}
\tablewidth{0pt}
\tablecaption{Characteristics of the Selected Regions}
\tablehead{
\colhead{} & \colhead{Region 1} & \colhead{Region 2} & \colhead{Region
             3} & \colhead{Region 4} & \colhead{Center} &
             \colhead{Total} & \colhead{Region A} & \colhead{Region B}} 
\startdata
RA (J2000) & 9:55:52.42 & 9:55:50.63 & 9:55:52.74 & 9:55:50.22 & 
9:55:51.19 & 9:55:51.32 & 9:55:52.17 & 9:55:52.68 \\
Dec (J2000) & +69:40:32.1 & +69:40:45.6 & +69:40:48.5 & +69:40:58.60 &
+69:40:46.8 & +69:40:45.30 & +69:41:12.5 & +69:40:22.2 \\
Size & $6\arcsec\!.8\times6\arcsec\!.8$ &
$6\arcsec\!.8\times6\arcsec\!.8$ & $6\arcsec\!.8\times6\arcsec\!.8$ &
$6\arcsec\!.8\times6\arcsec\!.8$ & $11\arcsec\!.3\times11\arcsec\!.3$
& $24\arcsec\!.9\times24\arcsec\!.9$ & $14\arcsec\!.8\times20\arcsec\!.3$ & $14\arcsec\!.8\times20\arcsec\!.3$  
\enddata
\label{tabregions}
\end{deluxetable*}

The observations were made with the Infrared Spectrograph (IRS)\footnote[8]{The IRS was a collaborative venture between Cornell University and Ball Aerospace Corporation funded by NASA through the Jet Propulsion Laborator and the Ames Research Center} spectrometer \citep{houck04}
on board the \textit{Spitzer Space Telescope}, under the IRS
guaranteed time observing program. The data were taken on June 6th,
2005 using the IRS ``mapping mode'' in all four modules: Short-High
(SH; $10 - 19\mu$m), Long-High (LH; $14 - 38\mu$m), provide $R\sim600$, while Short-Low (SL; $5 - 14\mu$m) and Long-Low (LL; $14 - 38\mu$m) give $R\sim60 - 130$. Each of the SL and LL modules are further divided into two subslits, which correspond to diffraction orders: SL1 ($7.5 - 14\mu$m), SL2 ($5 - 7.5\mu$m), LL1 ($20 - 38\mu$m), and LL2 ($14 - 20\mu$m). The SH map
consists of 30 pointings with 4 cycles each, and each subsequent pointing is
offset by half a slit width parallel to the slit and about one third
of the slit length along the slit.  The SH map covers an area of
$28\arcsec\times 23\arcsec$. The LH data consists of 12 pointings,
with 5 cycles each, covering an area of $38\arcsec\times
33\arcsec$. The offsets are equivalent to the SH map.  The SL data
consists of 120 pointings, with 2 cycles each, covering an area over
the M82 central region of $55\arcsec.5\times 57\arcsec$. The LL data
consists of 22 pointings with 2 cycles each.  Both SL and LL maps
follow the same offseting scheme as SH.  Because of the high
brightness of M82 it was unavoidable that the LL1  data became saturated
near the center of M82, but the LL2 data are still usable.  Fig.~\ref{figregions} (left) shows the areas
covered by the IRS SL, LH, and SH maps overlayed on the IRAC $8\mu$m
image from \citet{engel06}. The total integration times range from
12 (for SL exposures) to 31 sec (for LH exposures). Due to issues concerning the extraction, the LH spectra were not used in this analysis.
The boxes 'A' and 'B'
are the regions where the complete low-res (SL+LL) spectra were extracted.

The data were processed with version 13.2 of the Spitzer reduction
pipeline (version 14 for LL). Observations taken at each position were
combined into spectral cubes using CUBISM \citep{smith07}, 
an IDL-based software package designed to combine spectral mapping
datasets into 3D spectral cubes. Bad pixels in the basic calibrated data (BCD) spectra were manually flagged
and then automatically discarded when rebuilding the cube. Spectra
from off-source positions 1 kpc to the NE of the nucleus were used to subtract the background
from the low-resolution spectra.  For the high-resolution spectra we
did not subtract a background since there was no suitable ``sky''
spectrum available and the high source fluxes strongly dominate any background
emission.  The spectral analysis was done using SMART
\citep{higdon04} and PAHFIT \citep{smith06}. 

\section{Analysis}

On the basis of the SH spectral map coverage we defined six sub-regions
for which we extracted the spectra from SL and SH with CUBISM.  The
location of these regions is shown in Fig.~\ref{figregions} (right)
overlayed on the IRAC 8$\mu$m image from \citet{engel06}. The ISO aperture is also shown to illustrate the increase in spatial resolution obtained with $Spitzer$-IRS. Regions 2
and 3 coincide with the peaks of the [NeII] emission, regions 1 and 4
are offset to both sides of the galactic disk, the slightly larger
region C covers the nominal center of the galaxy, and the last region
corresponds to the entire area mapped with the SH spectrograph. The exact coordinates and
sizes are listed in Table~\ref{tabregions}.

\begin{figure*}
\epsscale{1.10}
\plottwo{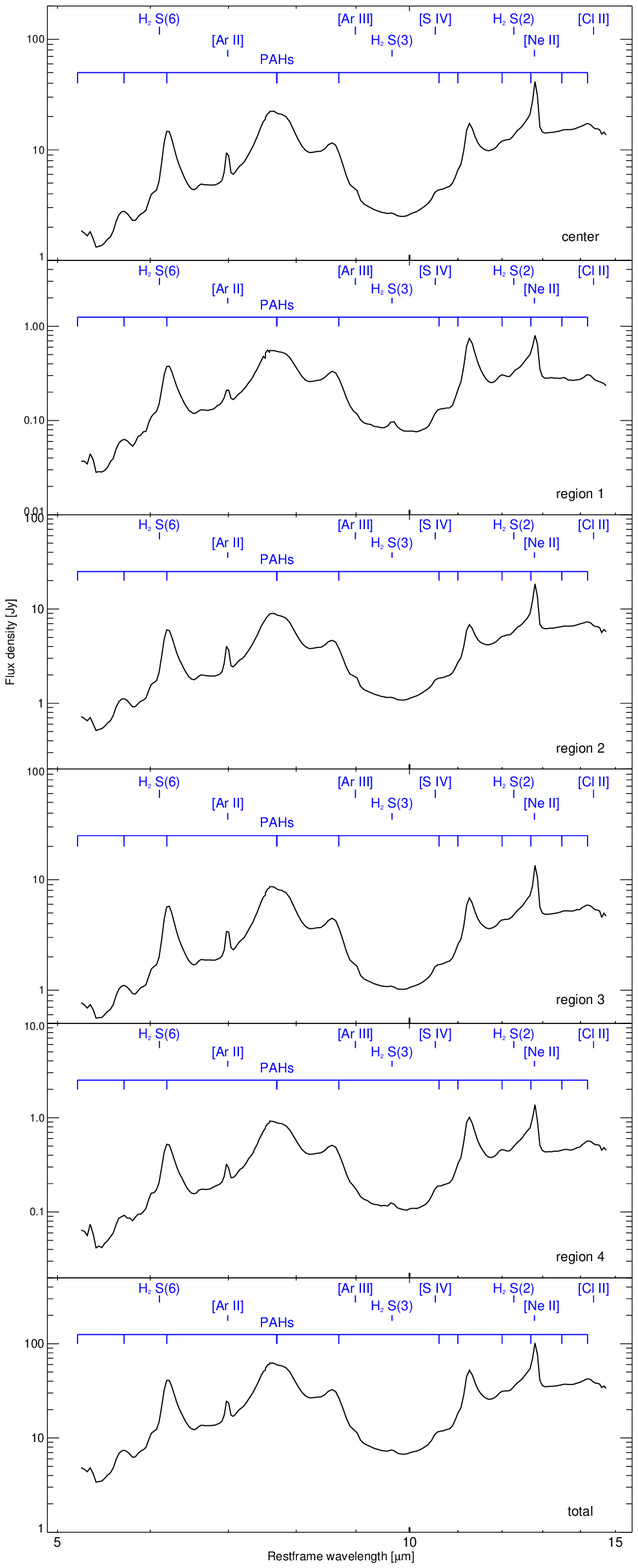}{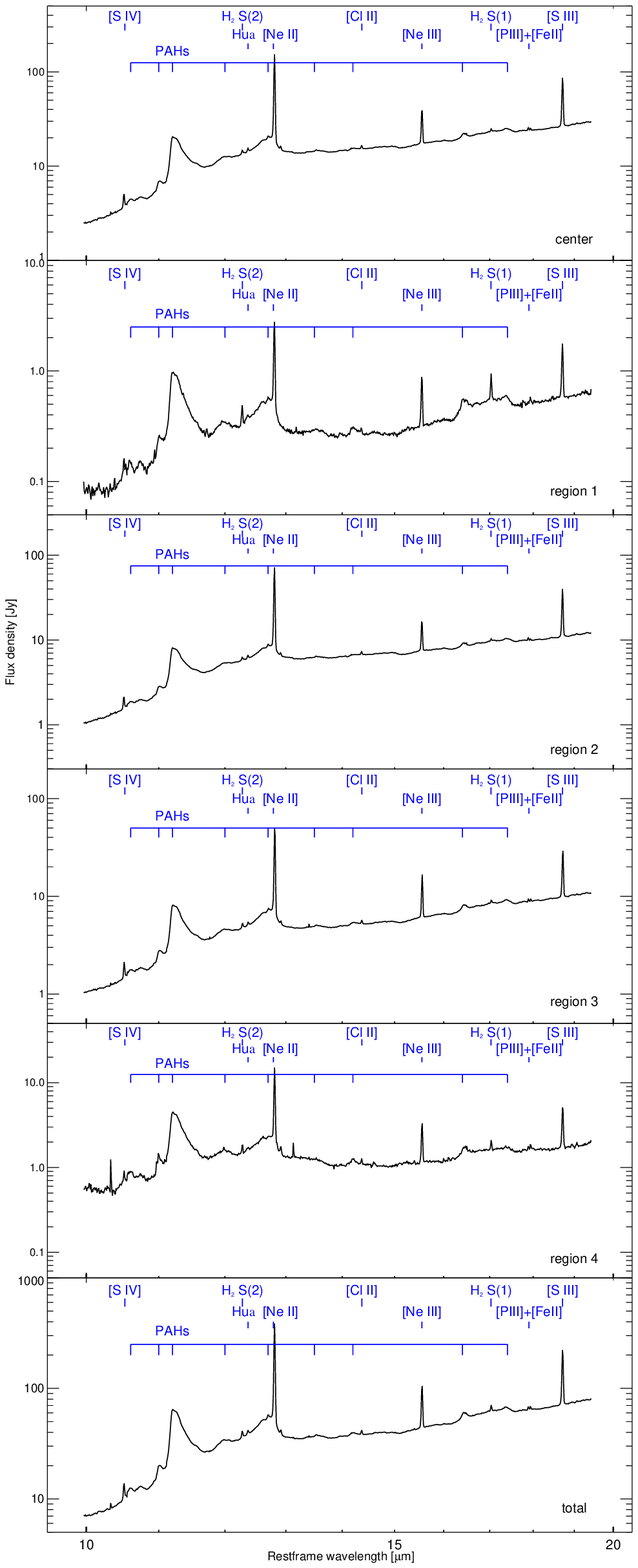}
\caption{SL (left) and SH (right) spectra of the six selected regions
         within the central starburst region of M82. The ``Total'' region corresponds to the total SH map in Fig.~\ref{figregions}.}
\label{figSLSH}
\end{figure*}

Fig.~\ref{figSLSH} shows the SL ($5 - 14\mu$m), and SH ($10 - 19\mu$m)
spectra extracted for these six regions. The noise is negligible and
the spectra exhibit the classical features of starburst galaxies
\citep{brandl06}, such as strong emission features of the PAHs, fine-structure lines, and
emission from molecular hydrogen, in addition to the broad silicate
absorption features around $9.7\mu$m and $18\mu$m, with an underlying continuum of very small grain emission (VSGs). The 5-38 $\mu$m
wavelength range contains many important diagnostic lines, such as
[ArII]$6.99\mu$m, [ArIII]$8.99\mu$m, [SIV]$10.51\mu$m,
[NeII]$12.81\mu$m, [NeIII]$15.56\mu$m, [SIII]$18.71\mu$m. All of the
detected features are labelled in Fig.~\ref{figSLSH}.  The fluxes and
ratios of the most relevant fine-structure and H$_2$ emission lines
are listed in Table~\ref{tablines}. These lines were measured using Gaussian fits to the line and linear fits to the local continuum.
The [NeII] line was measured
after subtracting the $12.6 \mu$m PAH feature. [SIV] at 10.5$\mu$m is hard to detect on regions 1 and 4 due to low S/N ratio, and [ArIII], at $8.99\mu$m is 
too close to the $8.6\mu$m PAH feature to be detected in a low resolution spectrum. Due to saturation of the LL1 module and extraction issues of the LH spectra, we could not measure the [SIII] $33.6\mu$m line.

\begin{figure}
\epsscale{1.2}
\plotone{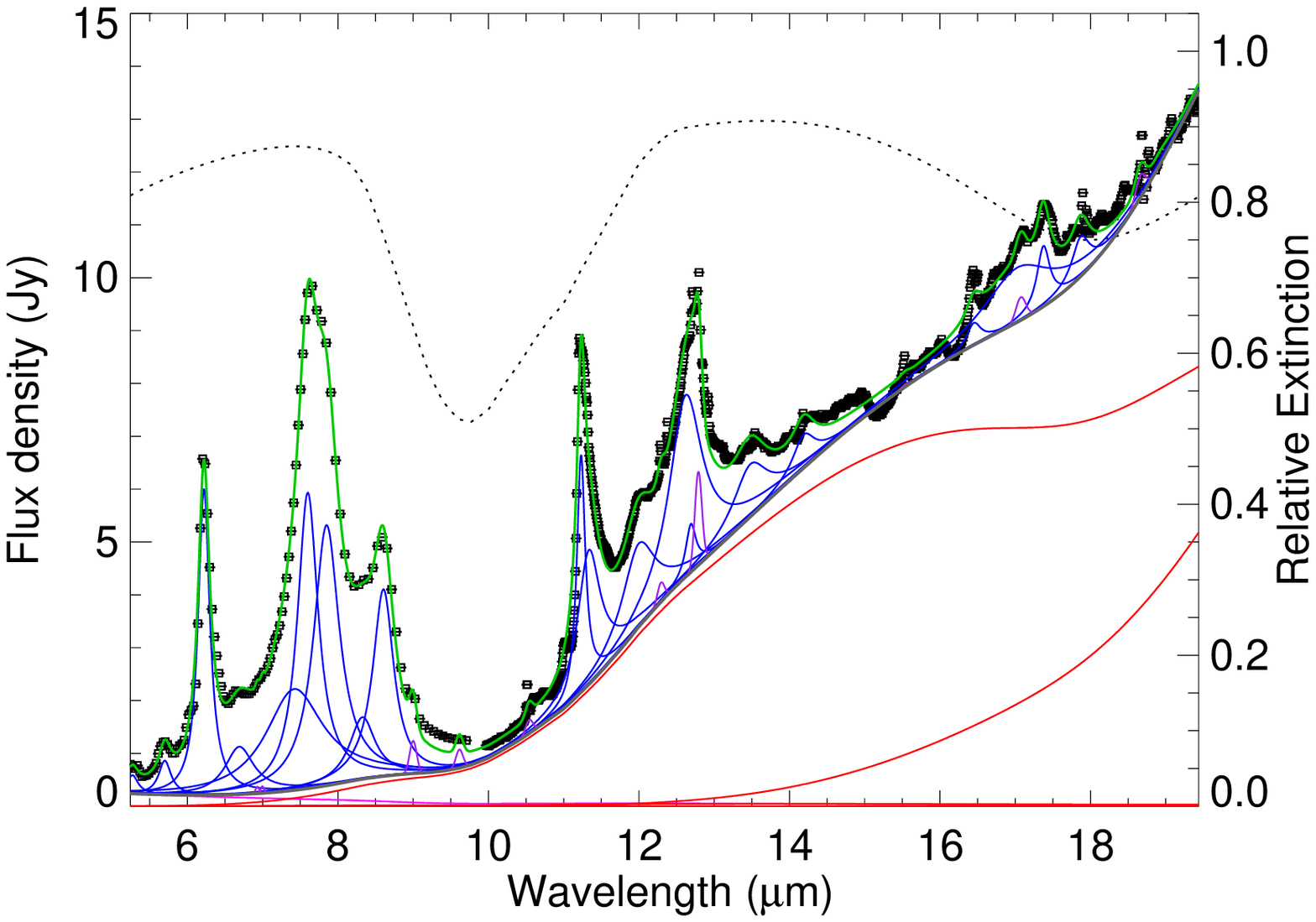}
\caption{Decomposition of SL+SH spectrum of Region 2. Red solid lines represent the thermal dust continuum components, the thick gray line the total continuum, blue lines are dust features, while the violet peaks are atomic and molecular spectral lines. The dotted black line indicates the fully mixed extinction which affects all components, with axis at right. The solid green line is the full fitted model, plotted on the observed flux intensities and uncertainties.}
\label{pahfit}
\end{figure}

Numerous PAH emission features are easily detectable in our spectra.
The fluxes and equivalent widths (EWs) of the strongest features at
$6.2 \mu$m, $7.7 \mu$m, $8.6 \mu$m, $11.3 \mu$m, $12.6 \mu$m,
$14.2 \mu$m, and the $16 - 18 \mu$m complex are listed in Table~\ref{tabpahs}.  Their values were
derived using PAHFIT \citep{smith06}, an IDL tool that decomposes low-resolution
spectra of PAH emission sources using a physically motivated
model. This model includes starlight, thermal dust continuum, resolved
dust features and feature blends, prominent emission lines, and dust
extinction. In our case, we merged SL and SH spectra in order to have one spectrum for each region with the widest wavelength coverage possible. Weaker 
PAH features at 5.2 $\mu$m, 5.6 $\mu$m, 12.0 $\mu$m, and 13.55 $\mu$m are present in the
spectra in Fig.~\ref{figSLSH}, but are not further utilised in
this paper. In Fig.~\ref{pahfit} we present an example of a fit to a combined SL+SH spectrum of region 2. In the overall spectrum (green) we distinguish 
the continuum component (red), PAH component (blue), ionic lines (purple), and dust absorption (dashed line). 

Silicate absorption affects mainly the wavelength interval from $7 - 12 \mu$m. In PAHFIT we have the option to include or exclude silicate absorption, and 
the difference between these two cases can be up to 50\% in flux of the $11.3 \mu$m feature. In section 4.3 we discuss the methods to estimate the magnitude and distribution of extinction.
Also there is a significant difference in the $11.3 \mu$m flux between the SH and SL spectra, which can also reach 50\%. This difference can be attributed 
to a poor fit to the silicate absorption feature in SH, as this module only covers wavelengths $>10\mu$m. In this paper we use the SH+SL measurements for 
this reason and to include the $17\mu$m complex. 

\begin{figure}
\epsscale{1.20}
\plotone{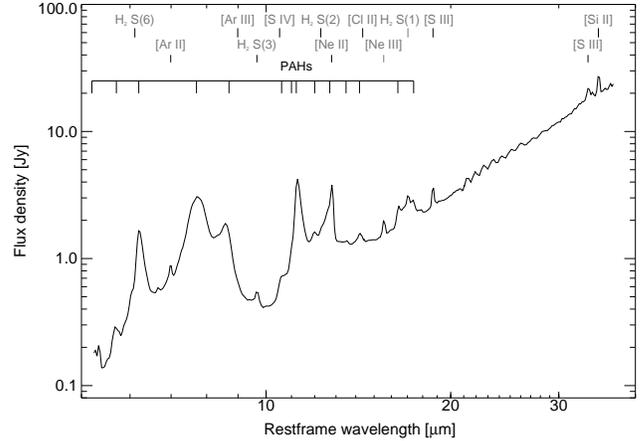}
\caption{Average of the $5 - 38\mu$m low-resolution IRS spectra of the regions A and
         B, located approximately 200\,pc above and below the galactic
         plane of M82.}
\label{figLL}
\end{figure}

We also extracted combined SL+LL spectra ($5 - 38\mu$m) from two $15\arcsec\times20\arcsec$ regions, A and B (see Fig. 1 left),  at a
distance of 200~pc above and below the galactic plane of M82, where
saturation did not compromise the LL measurements. Their positions are listed in Table 1. The spectra from regions A and B were virtually identical in shape, and an average of these spectra is shown in Fig.~\ref{figLL}. The spectrum is
clearly dominated by strong PAH emission features and a steeply rising
continuum, characteristic of classical starburst galaxies \citep[e.g.][]{brandl06}. The wiggles observed longwards $20\mu$m are fringes that originate from interferences in the detector substrate material.
At the low spatial resolution presented by the LL modules, we cannot see any unusual features in the SL+LL spectra of regions A and B. Also, as the LL1 slit is saturated at the central region of M82, we cannot extract any full low-res spectra of this region, connecting it to regions A and B. For these reasons, we will not make further analysis of LL1 module spectra in this paper.

\section{Results and Discussion}

The main gain of our observations over previous work on the starburst in M82 is the availability of spatially resolved mid-IR spectroscopy of the central region. Despite their overall similarity the spectra show distinct variations in the relative strengths of the spectral features, in particular in the neon fine structure lines and PAH features. These variations and their physical causes will be described in the following sections.

\subsection{The Morphology of the Starburst Region}

The discovery of a series of compact radio supernova remnants along
the galactic plane of M82, extending over 600~pc
\citep{kronberg85,muxlow94}, is an indication of very recent and
presumably ongoing star formation.  The detection of the ionic high
excitation lines [NeIII] and [SIV] confirms the
presence of very young massive stars in M82. Ratios using ionic lines of the same species and different ionization potentials such as [NeIII]/[NeII], [SIV]/[SIII], and [ArIII]/[ArII] are a useful measure of the
hardness and intensity of the radiation field and radiation density, and are therefore sensitive to the presence of young massive stars.  

Our measurements of the [NeIII]/[NeII] ratio are shown in Table~\ref{tablines}, for each of the six selected regions. We find
$0.13\leq$\,[NeIII]/[NeII]\,$\leq0.21$, with a median of [NeIII]/[NeII] = 0.18, which is consistent with the spatially
integrated ratio of [NeIII]/[NeII] = $0.16\pm0.04$ determined by
\citet{schreiber01}. These values are 30\% lower than the average value of 0.26
for the ISO-SWS sample of starburst galaxies \citep{thorn00}, and more
than an order of magnitude below the value of 8.5 found in the center of
NGC~5253 \citep{beirao06}, a low metallicity starburst galaxy at about the same distance as M82. 

Other ratios such as [SIV]/[SIII] and [ArIII]/[ArII] could be used to confirm the results on the [NeIII]/[NeII]. However, as mentioned in Chapter 3, our measurements of [SIV] in regions 1 and 4 have large errors from noise, and [ArIII], at $8.99\mu$m is too close to the $8.6\mu$m PAH feature to be measured accurately from the low resolution spectrum, precluding the use of these ratios. 

\begin{figure}
\epsscale{1.2}
\plotone{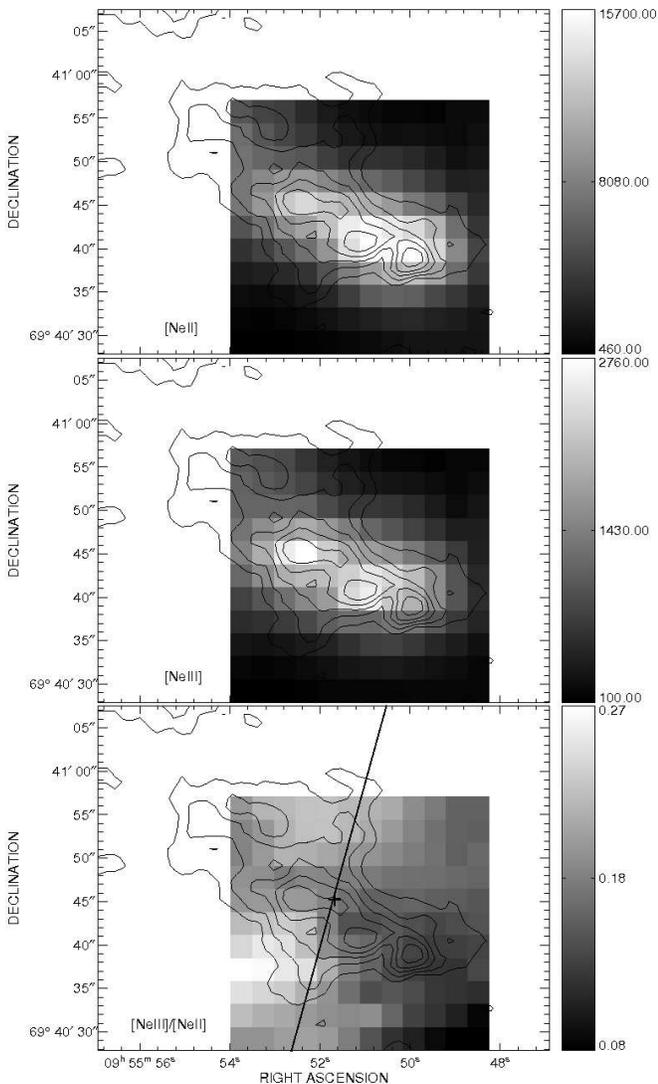}
\caption{Spectral maps in the [NeII] (top), [NeIII] (center), and
         [NeIII]/[NeII] (bottom) lines from the IRS SH module, with contour overlays of the $12\mu$m continuum emission from \citet{achter95}. The
         regions shown are $\sim26\arcsec\times26\arcsec$. At the bottom map, the line is the direction of the of the X-ray outflow, which is perpendicular to the plane of the galaxy. The cross represents the nucleus.}
\label{figlinemaps}
\end{figure}

The Spitzer/IRS spectral maps allow a study of the spatial variation of the radiation field, 
based on the [NeIII]/[NeII] ratio. Due to the presence of several luminous star clusters in the
central region, one might expect strong variations of [NeIII]/[NeII] between regions. 
In Fig.~\ref{figlinemaps} we present spectral maps of the two strongest neon
emission lines, and a ratio map, overlayed by the $12\mu$m continuum contours from \citet{achter95}. For both line maps, CUBISM was used to subtract a fitted continuum map from a total line+continuum map at the same wavelength on
a pixel-by-pixel basis. The [Ne II] emission in the upper map shows two peaks to either
side of the nucleus, a weaker and a stronger peak, which correspond to the E and W peaks in
\citet{achter95} respectively. After a close inspection, we identify the strong W peak in our maps with the two continuum emission peaks. The E and W [NeII] emission peaks are identified as a ``ring'' of ionized gas in \citet{achter95}. The morphology of the [NeII] emission follows the Br$\gamma$ emission in \citet{satyapal97} and \citet{schreiber01}. The [NeIII] map also reveals two emission peaks at the same positions, but with the eastern peak brighter relative to the western peak.

The lower map in Fig.~\ref{figlinemaps} presents the
[NeIII]/[NeII] ratio. The ratio varies from $0.08 - 0.27$ throughout the map. The lower value corresponds to the location of the westernmost clusters and there is a significant increase in the ratio further out from the galactic plane of
M82, from 0.15 to 0.27. Statistical $1\sigma$ errors are $\sim15\%$ and arise mostly from baseline determination errors. The grey line represents the direction of the X-ray outflow observed in M82 \citep{strick97,strick04}. It originates from the nucleus (marked with a plus sign) and is perpendicular to the plane of the galaxy. The peak of X-ray emission is offset from the outflow axis by 30 pc. Although the gradient in the [NeIII]/[NeII] ratio does follow the outflow axis, it appears offset by 5 arcsec to the east, and associated with the eastern cluster.

The overall morphology of the ionizing radiation in the region as revealed by the spectral maps seems to be in agreement with ground observations of the continuum and [NeII] emission by \citet{achter95}. However, the [NeIII]/[NeII] ratio varies only by a factor of three throughout the region and these variations do not correspond to the position of the emission peaks. An increase of the [NeIII]/[NeII] ratio is observed further out of the galactic plane, but due to the limited spatial coverage of the map, we cannot determine if this increase is connected to the outflows. The origins of the observed emission morphology and the variation of the [NeIII]/[NeII] ratio are discussed in the following subsection.

\subsection{Origins of the Variation of the Radiation Field}

The [NeIII]/[NeII] ratio measures the hardness of the radiation field, which is a function of stellar age and metallicity, and often parameterized by the effective temperature $T_{eff}$, and the radiation intensity as measured by the ionization parameter $U$\footnote{$U$ is defined as $U=\frac{Q}{4\pi R^2n_Hc}$, where $Q$ is the production rate of ionizing photons from the stars, $R$ is the distance between the ionizing cluster and the illuminated gas cloud, $n_H$ is the hydrogen number density of the gas, and $c$ is the speed of light.}.
Given the strength of the [NeII] and [NeIII] lines in M82, and their proximity in wavelength which minimises the effects of extinction, [NeIII]/[NeII] is the most reliable measure of the hardness of the radiation field. With the help of starburst models, it is possible to use the [NeIII]/[NeII] ratio to estimate the ages of the massive clusters in the region.  

Observations with ISO/SWS have been used previously for this purpose. \citet{schreiber01} determined a spatially integrated ratio of [NeIII]/[NeII] = 0.16 for the inner 500 pc of M82. Using the photoionization code CLOUDY and solar metallicity stellar atmosphere models by \citet{paul98}, \citet{schreiber01} modelled the variations of line ratios with $T_{eff}$. Adopting an electron density $n_e=300$ cm$^{-3}$ and an ionization parameter log$U=-2.3$, \citet{schreiber03} found for the ISO value for [NeIII]/[NeII] an effective temperature of 37400$\pm$400 K and a burst age of 4 -6 Myr. Other observed ratios were also modelled, such as [ArIII]/[ArII] and [SIV]/[SIII], giving similar $T_{eff}$ (within uncertainties). Independent estimations of cluster ages were done at longer wavelengths. \citet{colbert99} analysed far-infrared spectra from ISO/LWS and fitted line ratios to a combined HII region and PDR model. Their best fit model is an instantaneous starburst of 3 - 5 Myr old massive stars, in agreement with \citet{schreiber03}.

\begin{figure}
\epsscale{1.15}
\plotone{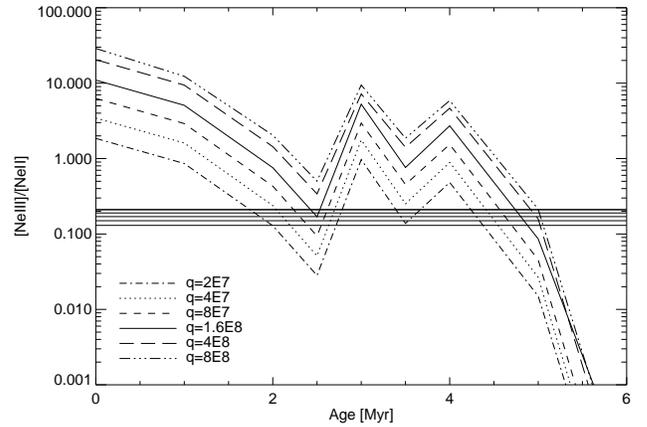}
\caption{Effect of the cluster age on the [NeIII]/[NeII] ratio. The data for our selected regions in M82 is shown by the horizontal lines. The model curves are computed for a cluster of $10^6M_{\odot}$, assuming Salpeter IMF, $M_{up}=100M_{\odot}$, and $M_{low}=0.2M_{\odot}$. Each curve represents a different ionization parameter, and the solid curve is the one that approaches the value found by \citet{schreiber01}, log$U=-2.3$. The horizontal lines indicate the [NeIII]/[NeII] for each selected region, from region 1 to region 4. The first dip in [NeIII]/[NeII] represents the ageing of the stellar population, after which WR stars are produced, increasing the [NeIII]/[NeII] ratio. The dip at 6 Myr occurs as most massive stars die through supernova explosions.}
\label{figmappings}
\end{figure}

If the infrared emission peaks correspond to massive clusters of stars, we can determine their ages from the measured [NeIII]/[NeII] ratio and compare them with the above results, using the photoionization models by \citet{snijders07}. As input, \citet{snijders07} used massive cluster spectra modeled with Starburst99, assuming a Salpeter IMF, $M_{up}=100M_{\odot}$, $M_{low}=0.2M_{\odot}$, and a gas density of $100$ cm$^{-3}$. This value of gas density is a factor of three lower than the \citet{schreiber01} value, but this has a small effect on the [NeIII]/[NeII] ratio. Fig.~\ref{figmappings} show the results for a range of ionization parameters. The selected regions are represented by horizontal lines. The value of log$U=-2.3$ derived by \citet{schreiber01} corresponds to a ionization front speed of $q=1.6\times10^8$cm s$^{-1}$. For a typical value of $q$, the ages of the clusters in each region range from 5 - 6 Myr, in agreement with the previously determined burst ages. These clusters dominate the central region of M82 and may be similar to those observed further out in M82 in the optical, studied in detail by \citet{smithlj06} using HST/ACS, which were found to have an average age of $6.4\pm0.5$ Myr.

As shown in Fig.~\ref{figlinemaps}, the [NeIII]/[NeII] ratio increases from 0.15 to 0.27 with increasing distance from the galactic plane of
M82. This is counterintuitive, as one might expect a harder radiation field at the location of the most luminous regions along the plane. Away from the plane, a decrease of gas density, relative to the number of ionizing photons, leads to an increase of the ionization parameter, which then causes an increase of the [NeIII]/[NeII] ratio, as discussed in \citet{thorn00}. Indeed, Fig.~\ref{figmappings} shows that the [NeIII]/[NeII] scales with the ionization parameter. The variation we observe in the [NeIII]/[NeII] ratio implies a variation of a factor of five in the ionization parameter. This is equivalent to saying that the gas density decreases five times faster than the radiation field which decreases as $\sim R^{-2}$ with $R$ being the distance to the ionization source. Shocks could also contribute to the increase of the [NeIII]/[NeII] ratio in the outflow region, but in Section 4.5 we show that to be minimal. This hypothesis could be tested using the [SIII]$18.6\mu$m/[SIII]$33.6\mu$m. Unfortunately, due to the problems reported in Sec. 2, we could not derive an accurate flux for the [SIII]$33.6\mu$m line in both LL and LH spectra. 

It is important to emphasize that, even at higher angular resolution, the
[NeIII]/[NeII] ratio in M82 remains quite low for an active starburst. We would have expected a larger variation with higher ratios locally corresponding to younger clusters and lower elsewhere.
A comparison with the ISO-SWS sample of starburst galaxies (\citet{thorn00}) shows that it is actually lower than most starbursts, despite being closer and better resolved. The low [NeIII]/[NeII] ratio could be caused by an aged stellar population where the starburst activity ceased more than half a dozen Myrs ago - although this possibility seems unlikely given the large amounts of molecular gas still present at the center of M82. An edge-on view of the galaxy could also contribute to the low variation of [NeIII]/[NeII] ratio.
 
As a comparison, we examined the $4\arcsec.5\times4\arcsec.5$ area with the highest [NeIII] flux (Fig.~\ref{figlinemaps}), comparing the measured [NeII] and [NeIII] luminosities with the models to determine the enclosed stellar mass. A single super star cluster of 5 Myr would have a cluster mass of $10^6M_{\odot}$ which is twice the mass of the super star cluster in NGC 5253 \citep{turner03}. A single cluster in a $4\arcsec.5\times4\arcsec.5$ area would correspond to a cluster number density of $\sim200$/kpc$^2$, which is comparable to the cluster density found in the fossil starburst region of M82 by \citet{degrijs01}. 

While the average age of the starburst population in M82 appears to be $\sim$5 Myr, ongoing star formation ($\le$1 Myr) could possibly be obscured by recent contributions of older stellar populations ($>5$ Myr) to the neon ratio. Considering this possibility, we explore the region with the highest [NeIII] flux, a strong indicator of the presence of O stars, as an illustrative case to set an upper limit on the ongoing star formation in M82.
This region has a [NeIII]/[NeII] of $\sim 0.10$, with a total [NeIII] luminosity of $1.41\times10^6L_{\odot}$. Assuming $\sim1.6\times10^8$cm s$^{-1}$, a typical young cluster of 1 Myr has [NeIII]/[NeII]$\sim5$ and can
contribute $\sim$50\% to the total [NeIII] luminosity or $\sim$2.5\% per mass
relative to the older 5$-$6 Myr old population. This means that the 1 Myr old population only emits $\sim$5\% of the total [NeII]+[NeIII] luminosity of the older 5$-$6 Myr old population. Even as an upper limit, this value indicates that the activity of the
starburst has substantially declined relative to the high star
formation rate that existed 5 Myr ago.

Despite the reduced starburst activity, the presence of CO emission all over the central region (Fig.~\ref{figsilicate}) shows that the starburst in M82 still has a large gas reservoir, as pointed out by \citet{thorn00} and \citet{schreiber03}. It appears therefore unlikely that the starburst activity ceased because of lack of fuel. On the other hand, negative feedback mainly through strong stellar winds and supernovae explosions can play an important role in determining the star formation rate in starbursts. \citet{schreiber03} calculated a feedback timescale of $1 -10$ Myr, which is in good agreement with the age of the older super star clusters in the central region of M82.

In summary we find that [NeIII]/[NeII] ratio is low on average, and increases with distance from the galactic plane of M82. The increase can be explained by an increase of the ionization parameter through a drop in gas density. The low [NeIII]/[NeII] ratio indicates that the dominant population consists of older clusters ($>5$ Myrs). We cannot rule out the presence of younger clusters in the central region, but at a much reduced rate ($<5\%$) of star formation compared to previous epochs. The large amount of molecular gas still present in the central region argues against a starvation of starburst activity, due to lack of gas. Instead, we believe that negative feedback processes are responsible for the observed decline in the star formation rate.

\subsection{Extinction}

One of the most noticeable features in the spectra in Fig.~\ref{figSLSH} is the absorption feature at $9 - 11\mu$m, caused by silicate grains. This feature can be extremely deep especially in ULIRGS \citep[e.g][]{spoon07}, and affects the observed fluxes of spectral lines and features in this region.
The strength of this feature is characterized by the optical depth at $9.8\mu$m ($\tau_{9.8}$) assuming a simple geometrical dust distribution. For our analysis it is important to estimate the intensity of this feature and its spatial variation in order to investigate its influence on the PAH strengths, which we discuss in the following subsections.

To study the distribution of $\tau_{9.8}$ in the SL region, we selected an area of $20\times12$ pixel in the center of the SL1 map. Within that area we extracted spectra from 60 spatial resolution elements ($2\times2$ pixel each).

\begin{figure}
\epsscale{1.15}
\plotone{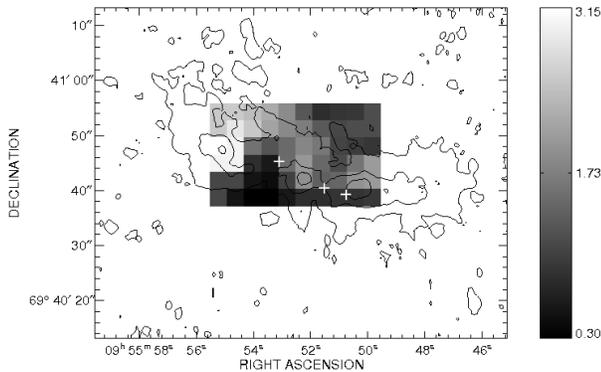}
\caption{Map of $\tau_{9.8}$, determined from a simple continuum fit to the SL spectra, with overlay of CO 1-0 emission contours from \citet{shen95}. The crosses indicate the [NeII] emission peaks. The image was rotated by 55 degrees and interpolated.}
\label{figsilicate}
\end{figure}

The apparent optical depth $\tau_{9.8}$ is then estimated from the ratio of the interpolated continuum to the observed flux at $9.8\mu$m. For the central region of M82, $\tau_{9.8}$ ranges from 0.3 - 3.1, assuming a foreground screen attenuation. 
We built an extinction map of $10\times6$ resolution elements from the simple fit method, which is shown in Fig.~\ref{figsilicate}. The qualitative distribution of $\tau_{9.8}$ observed in this figure is very similar to the qualitative results from the PAHFIT fitting method. There is a good correlation between the $\tau_{9.8}$ distribution and the CO 1-0 emission, indicating that dust and molecular gas coincide in this region. The enhanced $\tau_{9.8}$ in the northwest region of the map indicates an increase of silicate dust above the galactic plane. 

For an independent, and possibly more accurate, estimate of $\tau_{9.8}$, we used PAHFIT on 15 SL+LL1 spectra ($5 - 20\mu$m), each corresponding to 4 resolution elements fitted only with SL. With the PAHFIT method, $\tau_{9.8}$ ranges from 0 - 2.52, with a median is 1.34. 

The combined SL+LL spectrum is necessary to better constrain the parameters in PAHFIT. Fitting only SL spectra with PAHFIT can result in large errors in the calculation of $\tau_{9.8}$. Unfortunately, due to saturation at wavelengths longward of $20\mu$m, we have an insufficient spatial coverage of LL data in the central region. In addition, the LL slits are $\sim10\arcsec.$ wide, providing low spatial resolution. 

For the higher spatial resolution, we used the simple method described by \citet{spoon07} to estimate $\tau_{9.8}$. In this method we approximate the mid-IR continuum at $9.8\mu$m by a power law fit to the flux pivots at $5.5\mu$m and $14.5\mu$m, avoiding the PAH emission features.

The values derived from both SL+LL and simple fit methods agree qualitatively well, but in regions where $\tau_{9.8}<1$, the difference between the two methods is greater than a factor of two. Using a mixed attenuation law with $A_v/\tau_{9.8}=16.6$ \citep{rieke85}, the range of $\tau_{9.8}$ corresponds to $0<A_v<41.8$ for the SL+LL method. The values of $\tau_{9.8}$ from the power-law interpolation method give $5.0<A_v<51.5$. These values are in agreement with \citet{schreiber01}, which derive $23<A_v<45$ for their selected regions in M82, which cover an area closer to the infrared peaks.  

We conclude that both methods, besides their significant uncertainties in the magnitude of $\tau_{9.8}$, have consistently revealed significant variations in the amount of dust extinction across the central region. These variations are strong enough
to affect the following interpretation of PAH features.

\subsection{Variations of PAH Emission Features}

Polycyclic Aromatic Hydrocarbons (PAHs) are thought to be responsible
for a series of broad emission features that dominate the mid-infrared spectra of starbursts \citep[e.g.][]{peeters04}. They are
observed in a diverse range of sources
with their strongest emission originating in photodissociation
regions (PDRs), the interfaces between {\sc HII} regions and molecular
clouds.

The relative strength of the different PAH bands is expected to vary with the size and the ionization state of the
PAH molecule \citep{draine01,draine06}. Observations of Galactic sources
\citep[e.g.][]{verst96,joblin96,vermeij02} have shown that the relative strengths of individual PAH
features depend upon the degree of ionization: C-C stretching modes at
$6.2\mu$m and $7.7\mu$m are predominantly emitted by PAH cations, while the C-H
out-of-plane bending mode at $11.3\mu$m arises mainly from
neutral PAHs \citep{draine01}. Thus the ratios $6.2/11.3\mu$m and $7.7/11.3\mu$m may be used as indicators of PAH ionization state. \citet{joblin96} have found that the 8.6/11.3 $\mu$m ratio can also be linked to variations in the charge state of the
emitting PAHs, which is supported by laboratory experiments \citep{hudgins95}. \citet{smith06} found band strength variations 
of factors of 2--5 among normal galaxies.

\begin{figure}
\epsscale{1.3}
\plotone{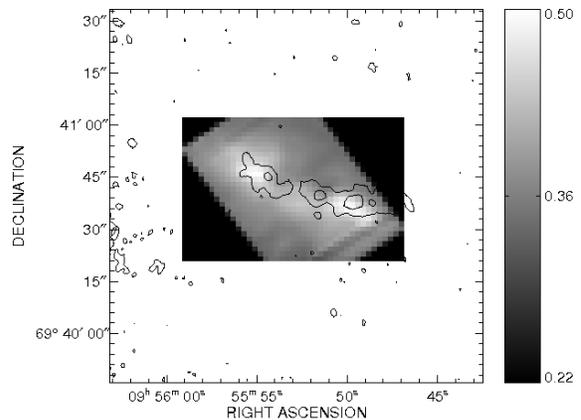}
\caption{IRS spectral map of the PAH ratio $6.2/11.3\mu$m, with overlay of CO 1-0 emission contours from \citet{shen95}. The map is in 
         logarithmic scaling. This map was done using CUBISM maps of baseline-subtracted flux at 
$6.0 - 6.5\mu$m and $11 - 11.7\mu$m. 
         }
\label{figpahmap}
\end{figure}

Studies of PAHs in M82 have been done previously using ISO. Observations with ISOCAM \citep{schreiber02} revealed a decrease in the $6.2/7.7\mu$m ratio 
and an increase in the $8.6/11.3\mu$m PAH ratio from the nucleus outwards along the galactic plane. The observations also showed a good spatial correlation 
of the $8.6/11.3\mu$m ratio with the CO (1-0) emission. These ratio variations are attributed to real differences in the variation of physical characteristics 
of PAHs across M82, specifically a higher degree of PAH ionization within the most intense starburst sites. 

We use CUBISM to build maps of ratios of PAH features that exist in the 6 - 14 $\mu$m SL spectra. The $7.7\mu$m feature is 
split between the two SL orders, and the $8.6\mu$m feature is largely influenced by the $7.7\mu$m feature, making the local continuum determination difficult. 
For these reasons the $6.2\mu$m feature map was chosen to be compared to the $11.3\mu$m feature map. In Fig.~\ref{figpahmap} we present the 
$6.2/11.3\mu$m PAH ratio map, overlayed with the CO (1-0) contours from \citet{shen95}. 

Obviously, the distribution of the  $6.2/11.3\mu$m PAH ratio correlates well with the molecular ring defined by the CO (1-0) map. The molecular emission 
forms two lobes $\sim200$ pc from the nucleus whereas the ionized emission is
concentrated around the nucleus, in the regions where the [NeII] and [NeIII] emissions peak. 

However, the $6.2/11.3\mu$m PAH ratio may be affected by extinction.
Comparing the CO emission with the extinction map in Fig.~\ref{figsilicate} and with the PAH ratio map we see that they correlate well. Hence, we investigate the effects of extinction on the observed PAH ratios in the next subsection.

\begin{figure}
\epsscale{1.2}
\plotone{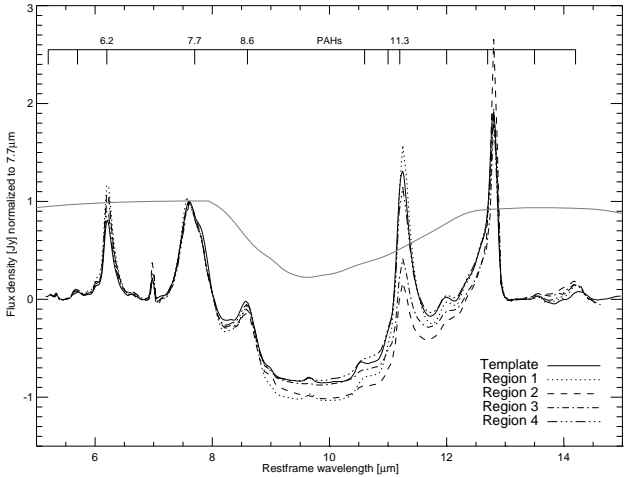}
\caption{Baseline subtracted IRS spectra of the selected regions from Fig. 1. The gray line is the \citet{chiar2005} extinction law, applied to a normalized flux of 1.75, and assuming $\tau_{9.8}=2$.}
\label{figpahplot}
\end{figure}

\subsubsection{Influence of silicate absorption on PAH ratios}

The fluxes of the PAH bands at $6.2\mu$m, $7.7\mu$m, $8.6\mu$m, and $11.3\mu$m are all affected by dust absorption. However, as the $11.3\mu$m is significantly more 
affected by the $9.8\mu$m silicate feature than the $6.2\mu$m feature, this will affect the measured PAH ratio in 
Fig.~\ref{figpahmap}. To illustrate this influence, we show in Fig.~\ref{figpahplot} the baseline subtracted spectra of the selected regions around the 
nucleus in M82, normalized to the flux at $7.7\mu$m to emphasize the relative flux variations of the $11.3\mu$m PAH feature. The gray line is the 
\citet{chiar2005} extinction law, applied to a normalized flux of 1.75, and assuming $\tau_{9.8}=2$. For this figure, the baseline was removed by 
subtracting a second order polynomial fitted to the following wavelengths: $5.5\mu$m, $6.8\mu$m, $8.0\mu$m, $13.2\mu$m, and $14.5\mu$m. These wavelengths 
were chosen to avoid the silicate feature at $10\mu$m and the PAH features. 
The effect of extinction on the $8.6\mu$m and $11.3\mu$m PAH features is similar, meaning that the $8.6/11.3\mu$m ratio could possibly be used to study PAH 
ionization with minimal concern for extinction, as the difference in flux correction between $8.6\mu$m and $11.3\mu$m is $\sim15\%$.
However, the $8.6\mu$m feature is influenced by the broad PAH feature at $7.7\mu$m, which makes the local continuum fitting more difficult compared to the 
$6.2\mu$m feature. 

\begin{figure}
\epsscale{1.2}
\plotone{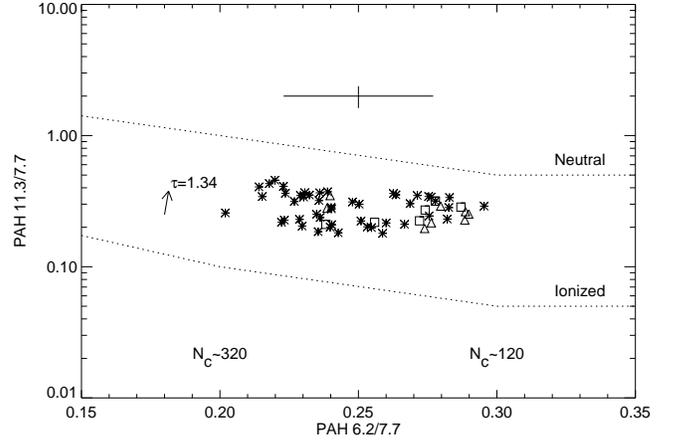}
\caption{The $6.2/7.7\mu$m PAH ratio vs. $11.3/7.7$ PAH ratio, for the same sub-regions as in the absorption map in Fig.~\ref{figsilicate}. The triangles are points from the areas where [NeIII]/[NeII]$>0.22$ and the squares are points from the areas where [NeIII]/[NeII]$<0.13$. Error bars represent average uncertainties, related to extinction correction and fitting uncertainties. The arrow represents the effect of a silicate absorption feature with $\tau_{9.8}=1.34$. The tracks represent the ratios of a population of neutral and ionized PAHs.}
\label{figdraine}
\end{figure}

Fig.~\ref{figpahplot} suggests that the variations of PAH feature ratios involving the $11.3\mu$m feature will be heavily affected by extinction and contributing to the distribution of the $6.2/11.3\mu$m PAH ratio in Fig.~\ref{figpahmap}. 
The $7.7\mu$m PAH feature is affected by extinction to a similar level as the $6.2\mu$m feature, as shown in the Fig.~\ref{figpahplot}. \citet{draine01} use the $6.2/7.7\mu$m ratio for PAH size diagnostic and $11.3/7.7\mu$m for a PAH ionization state diagnostic. Variations of these ratios reflect real variations of the physical properties of PAHs in M82 only if their variations are not due to extinction effects. Fig.~\ref{figdraine} presents the $6.2/7.7\mu$m and $11.3/7.7\mu$m PAH ratios calculated from PAHFIT measurements of the same sub-regions as in the extinction map. The data in Fig.~\ref{figdraine} is corrected for extinction. We took the average of the extinction methods (continuum fit method and the PAHFIT fit on SL+LL1 spectra (see 4.3)) for correction. The points are dispersed between the tracks representing totally ionized and neutral PAH populations, and the error bars show the average of the difference between the extinctions derived by the two methods. The triangles are points from the areas where [NeIII]/[NeII]$>0.24$ in the [NeIII]/[NeII] map from Fig.~\ref{figlinemaps}. These are regions with harder radiation field, where a greater number of ionized PAH are expected to be observed. The squares are points from areas in the map where [NeIII]/[NeII]$<0.13$. To preserve the clarity of the plot, only the error bars at these points are represented, as they are typical for all the points.
The arrow represents the effect on the ratios of a silicate absorption feature with $\tau_{9.8}=1.34$, which is the median optical depth for the region as explained in section 4.3. 

The $11.3/7.7\mu$m PAH ratios are dispersed between 0.2 and 0.5.  However, the error bar can account for this dispersion, demonstrating how silicate absorption can affect the diagnostic of PAH ionization state based on this ratio, on scales of the size of the 
$IRS$ resolution elements, $\sim35$ pc. This implies that the average variations in the ionization state of the grains are relatively small on scales of 35 pc. These results are to be compared to SINGS results, in which the $11.3/7.7\mu$m PAH ratio is found within the same 
range as in the central region of M82 for galaxies dominated by HII regions \citep{smith06}.

The $6.2/7.7\mu$m ratio varies between $\sim$0.2 and 
$\sim$0.3. This is well in the range of $0.2 - 0.4$ reported by \citet{smith06} for SINGS galaxies dominated by HII regions. We observe no significant difference in the $6.2/7.7\mu$m ratio between regions with high [NeIII]/[NeII] ratio and regions with less [NeIII]/[NeII] ratio. Extinction and fitting errors affect the $6.2/7.7\mu$m ratio only by $\sim1\%$, and errors from fitting residuals from PAHFIT amount to less than 2\%. However, there are significant uncertainties arising from the continuum fitting by blackbody curves with effective temperatures between $35 - 300$K. Modifying the number of blackbody components and their temperatures results in changes of the PAH ratios in excess of $8 - 12\%$. The horizontal error bar represents the average uncertainty of $10\%$.

The range of $6.2/7.7\mu$m ratios implies an environment composed by a warm ionized medium and photodissociation regions (PDRs) \citep{draine01}. The dispersion in the data could reflect a real variation of this ratio. The general significance of the 
variation of the $6.2/7.7\mu$m is discussed on \citet{draine01}. Assumptions about the stellar radiation intensity, which affect the $11.3/7.7\mu$m ratio as well, account for these variations in the number of C atoms,  Following \citep{draine01}, our observed ratios correspond 
to a number of carbon atoms in a PAH grain between ~100-140 ($6.2/7.7\mu$m$\sim0.3$) and ~240-320 ($6.2/7.7\mu$m$\sim0.2$).

However, there is no correlation between the $6.2/7.7\mu$m ratio with radiation hardness (symbols in Fig.~\ref{figdraine}) or with any other resolved
spatial structure in the central region. Since the variations are comparable to the uncertainties in the measurement the results are not (yet) significant
enough to support PAH size variations at parsec scales.
Similarly, the variations of the $11.3/7.7\mu$m ratio are mostly due to extinction and show little support for variations in PAH ionization throughout the 
region. However, if PAH sizes would vary on parsec scales one would expect a strong emission from the larger grains which radiate predominantly at larger 
wavelengths, such as the $17/mu$m complex.

\subsubsection{The $17 \mu$m PAH complex}

The $16 - 18 \mu$m wavelength range contains the H$_{2}$ S(1)17.03 $\mu$m line and
the $17\mu$m PAH complex. This complex is attributed
to a blend of emission features ($16.45\mu$m, $17.03\mu$m, and $17.37\mu$m) which are possibly due to PAH C--C--C
bending modes \citep[e.g.][]{kerck00}, and also emission from PAH clusters, amorphous carbon particles, and other PAH-related species \citep{peeters04a}. Following the \citet{draine06} models, the $17\mu$m complex is mostly emitted by large PAH molecules with 1000 - 2000 carbon atoms, while the $6.2\mu$m feature is emitted mostly by smaller PAH molecules with only 200 - 300 carbon atoms. 
The relatively narrow features at $16.45\mu$m and $17.37\mu$m have been detected by ISO in Galactic \citep{moutou00} and extragalactic sources
\citep{sturm00}. Only with Spitzer-IRS has the whole $17\mu$m PAH complex been identified and routinely detected in both
normal and starburst galaxies \citep[e.g.][]{smith04,dale06,brandl06}, including the outer regions
of M82 \citep{engel06}. However, the spatial variation within galaxies other than the Milky Way has not yet been studied. \citet{engel06} detected this complex in 
the disk and in the halo of M82, but their spectral coverage did not include the central regions. Here we report the characteristics of this complex in the 
central kpc of M82.

As seen in Table~\ref{tabpahs} the ratios
between the $17\mu$m complex and the $11.3\mu$m PAH feature, as measured with PAHFIT, range
from 0.39 in regions 1 and 3 to 0.49 in the center region, and do not show clear spatial correlations. However, the $17/6.2\mu$m varies from 0.21 in 
region 3 to 0.57 in region 4. The ratios between the regions in the plane and the 
outward regions vary by a factor of 2. The latter numbers suggest a significant variation possibly due to different PAH sizes. However, correcting for an 
optical depth of $\tau_{9.8}=2$ for the regions 
around the [NeII] peaks (Fig.~\ref{figsilicate}), the ratio increases by $\sim67\%$, yielding values of 0.37 for region 2 and 0.35 for region 3. While 
substantially increased, the values are still more than 20\% below the values for region 1 and 4. However, due to uncertainties on the determination of $\tau_{9.8}$ we cannot diagnose any variation of PAH sizes based on this difference.

\begin{figure}
\epsscale{1.2}
\plotone{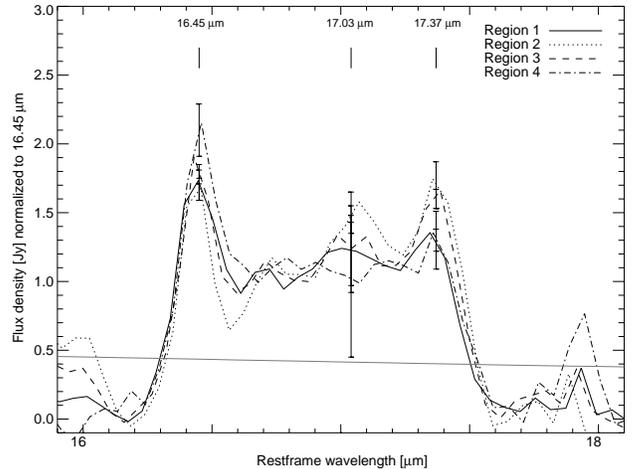}
\caption{Zoom-in on the PAH dominated emission features in the $16-18
         \mu$m spectral range of the SH spectra. The spectra were normalized to the average flux between $16.2\mu$m and $17.6\mu$m and rebinned to a resolution $R=300$, to diminish the noise. The error bars indicate 
noise at three wavelengths.}
\label{fig1618}
\end{figure}

Fig.~\ref{fig1618} shows a baseline-subtracted spectrum of the $17\mu$m PAH complex normalized to the average flux between $16.2\mu$m and $17.6\mu$m and 
rebinned to $R=300$, to 
enhance the broad components. In this figure we can see the flux variation in these three main PAH features between the regions. The differences in the 
strength of these features are consistent within the uncertainties, with exceptions being region 4 at 
$16.45\mu$m and regions 2 and 3 at $17.37\mu$m. Moreover, these discrepancies could be explained by errors in baseline subtraction, 
which are in the order of 10\%. 
The $17\mu$m PAH complex is also seen in
the LL spectra as far as 2 kpc from the galactic plane. The model of \citet{draine06} predicts that the relative strength of the individual components is
not sensitive to ionization state or grain size. Our observations are in good agreement with an invariant shape of this PAH complex.

\subsubsection{The difference between PAH emission and VSG emission}

Very small grains (VSGs) are dust particles larger on average than PAHs, with typical sizes in the range of 1 - 150 nm \citep{desert90}. They are excited by stochastic heating and are thought to be responsible for most of the mid-infrared continuum emission. The properties of these grains with relation to PAHs have been studied previously in galactic sources \citep{verst96,lebout07} and in dwarf galaxies \citep{madden06,wu06}. These studies focus on the behavior of VSGs in conditions where PAHs are destroyed, mainly by intense stellar radiation in HII regions or a low metallicity environment. Flux differences between VSGs and PAHs are reported in Galactic HII regions by \citep[e.g.][]{lebout07}, where the PAH/VSG emission ratio increases with the distance from the cluster.

As shown in Table 3, there are significant differences in EWs of the PAH features between the regions in the galactic plane and the regions outside the galactic plane. However, the EW shortwards of $10\mu$m behaves differently from the EW longward of $10\mu$m. Longwards of $10\mu$m, we observe that the EWs of the PAH features increase outwards the galactic plane. The $11.3\mu$m EW, for example, increases from 0.660$\mu$m in region 2 to 2.27$\mu$m in region 1. The EWs of the $6.2\mu$m and $7.7\mu$m decrease outwards the galactic plane. For example, the $6.2\mu$m EW decreases from 1.10$\mu$m in region 2 to 0.308$\mu$m in region 1. The EW of the $8.6\mu$m feature decreases up to a factor of two, from 0.865$\mu$m in region 3 to 0.437$\mu$m in region 2. This could be due to the contribution of the stellar continuum to the local continuum shortwards of $10\mu$m. The stellar continuum contribution decreases with wavelength and with the distance from the galactic plane. 

The continuum longwards of $10\mu$m is composed by thermal emission from VSGs.
As the PAH flux does not increase outwards from the galactic plane, this can only be due to a decrease of the local continuum emission relative to the PAH strength. This decrease can be explained by several different factors: photo-destruction, abundance differences between VSGs and PAHs, and different heating opacities between VSGs and PAHs with a varying radiation field. 

Decreasing PAH flux with the hardening of the radiation has been observed in galactic star forming regions \citep{verst96,lebout07}, on small spatial scales near the luminous clusters ($\sim$2 pc), where the PAH destruction largely surpasses PAH excitation. Wether the conditions that lead to PAH destruction on 2 pc scales can be maintained over much larger scales, corresponding to the resolution of our maps ($\sim$35 pc), cannot be derived from our data, but has been observed in NGC 5253 \citep{beirao06}.
Using the SINGS sample of galaxies, \citet{draine07} have shown that the fraction of PAH abundance over the total dust abundance decreases with metallicity. Hence, abundance differences between PAHs and VSGs could be observed in cases of a strong metallicity gradient. M82 has a metallicity gradient, but it becomes noticeable only at distances larger than 1 kpc from the center \citep{ranalli06}, which is greater than the distance from our regions 1 and 4 to the galactic plane (400 pc).

Our favored explanation is the difference in excitation between PAHs and VSGs, enhanced by variations of the radiation field. The mid-infrared PAH features are produced by vibrational-rotational modes of the PAH molecules, while the VSG continuum emission is mainly produced by thermal radiation. VSGs are bigger than PAHs, so thermal radiation becomes dominant over vibrational-rotational transitions. Following \citet{draine01} models, the opacity of the cross-section peaks at FUV wavelengths, where hot dust is needed to emit at $10 - 20\mu$m. However, at the distance from the galactic plane of regions 1 and 4, the radiation field is still intense enough to excite PAHs, but no longer of a high enough intensity to excite the dust to the same temperatures as in the plane of the galaxy. This provokes the PAH EW enhancement observed in these regions.



\subsection{Excitation of the Warm H$_{2}$}

Molecular hydrogen is the most abundant molecule in the Universe and can be used to probe the properties of the warm
molecular gas in M82. It can be traced in the mid-infrared through rotational emission lines, which may arise through three different mechanisms:
UV excitation in PDRs surrounding or adjacent to the {\sc HII} regions; shocks that accelerate and modify the gas in a cloud, collisionally exciting the 
H$_{2}$ molecules; hard X-ray photons capable of penetrating the molecular clouds and heating large ionizing columns of gas.

Vibrational-rotational H$_{2}$ line emission in M82 was studied in the near-infrared by \citet{pak2004}. By correlating the emission flux from these lines 
with [CII] $157\mu$m and far-infrared luminosity, they showed that the H$_{2}$ emission comes mainly from the PDRs. ISO observations of M82 detected S(0), 
S(1), S(2), S(6), and S(7) rotational lines \citep{rigo02}, excited by UV radiation from massive stars.

The S(1) and S(2) rotational transition lines of H$_2$ are clearly
detected in our SH spectra. The fluxes and temperatures of these lines are listed in
Table~\ref{tablines}, as well as the S(1)/S(2) ratios. The temperatures were calculated from the S(1)/S(2) ratios using the
method described by \citet{roussel06}, assuming an ortho- to para-
ratio of three. Our derived temperatures are in agreement with the
average temperature value of 450\,K, derived with ISO
\citep{rigo02}. The H$_2$\,S(0) line could not be detected in any of
the LH spectra, as its equivalent width is very low, supporting our finding that the H$_2$ temperature is indeed relatively high.

\begin{figure}
\epsscale{1.2}
\plotone{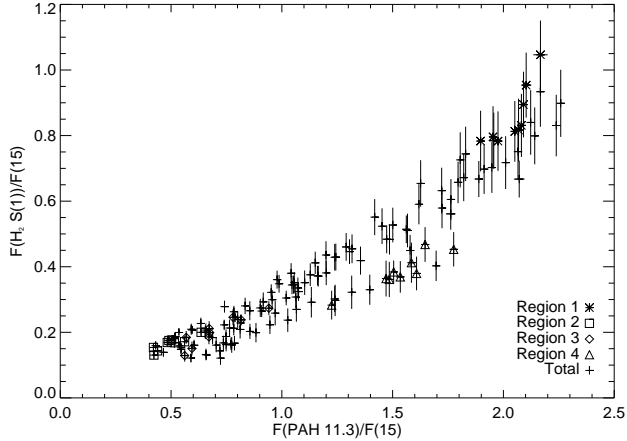}
\caption{The PAH $11.3\mu$m feature over the $15\mu$m continuum versus H$_2$S(1) over the $15\mu$m continuum, based on a pixel-by-pixel correlation between SH maps. The values corresponding to each of the selected regions are represented by different symbols. The error bars represent line flux measurement errors.}
\label{figh2corr}
\end{figure}

If H$_{2}$ emission is excited mainly by UV radiation in PDRs, the fluxes of H$_2$ lines and PAH lines should correlate closely, since PAH emission features 
arises from the same mechanism in PDRs. To look for secondary effects, like shocks, we plot in Fig.~\ref{figh2corr} the PAH $11.3\mu$m vs. H$_{2}$
S(1)$17.03\mu$m fluxes, both divided by the continuum flux at $14.8 - 15.2\mu$m to reduce the range covered by the figure. Each data point corresponds to a resolution element of the CUBISM flux maps taken from SH spectra. The stars correspond to region 1, squares to region 2, diamonds to region 3, and triangles to region 4. The error bars 
represent the line uncertainties from the measurement.  The correlation in Fig.~\ref{figh2corr} shows that the excitation mechanisms for
both species coincide at least on spatial scales of the resolution of the map, which is 35~pc.
While there is wider dispersion of data points corresponding to regions 1 and 4 (the regions associated with outflows) the systematic measurement 
errors account largely for this dispersion. We conclude that shocks in the outflow regions have no measurable influence on the H$_2$ emission. However, we cannot 
distinguish between UV- and local
shock excitation as produced by supernovae or energetic outflows, on scales smaller than the resolution of our pixels (35 pc).

\section{Conclusion}

We presented spatially resolved mid-infrared spectra of
the central region of M82.  The spectra were taken with the Spitzer
Infrared Spectrograph in both the $5-38\mu$m low-resolution ($R\sim
65-130$) and the $10-37\mu$m high-resolution $R\sim 600$ module. The
high signal-to-noise and the continuous spatial and spectral coverage
allowed us to study the nucleus of a starburst galaxy in unsurpassed
detail. Our goal was to study the physical conditions of the interstellar medium and their spatial
variations within the central kpc of M82.

Overall, the spectra show the typical features of a starburst:
prominent PAH features, silicate absorption, fine-structure lines, and
a steeply rising continuum. We built a spectral map with high-resolution spectra, selected six representative regions for spectral extraction, 
and studied the variations of the neon ionic lines and PAH feature emission among the regions.

We attempted to trace the structure of the ionizing radiation, and therefore the young stellar population of M82 through the diagnostic emission lines, [NeII]12.8$\mu$m, and [NeIII]15.5$\mu$m. The overall morphology of the ionizing radiation in the region as revealed by the spectral maps appears to be in agreement with previous ground-based observations \citep{achter95}. There is surprisingly little spatial variation of the [NeIII]/[NeII] ratio across the disk, and it varies only by a factor of three throughout the central region, with these variations not corresponding to the position of the emission peaks. We observed an increase of the [NeIII]/[NeII] ratio by a factor of 2 outwards the galactic plane,
which may be associated with the outflows. We suggest that the increase of the [NeIII]/[NeII] ratio with distance to the galactic plane is due to
a decrease in gas density rather than a hardening of the field.
 
The [NeIII]/[NeII] ratio is low on average, which indicates that the dominant population consists of 
already evolved clusters ($>5$ Myrs). We cannot rule out the presence of ongoing star formation in the central region, but it must occur at a much reduced rate ($<5\%$) compared to previous epochs. This drop is unlikely to be caused by starvation, as there is still a large amount of molecular gas present in the central region. It is more likely due to negative feedback processes causing a decrease in the star formation rate.

There are significant variations in the amount of dust extinction, which strongly correlate with the CO 1-0 emission across the central region. 
These variations are strong enough to affect the interpretation of PAH features, but due to limited spectral coverage the extinction estimates are uncertain.

The flux of the main PAH features correlates spatially with the flux of the neon ionic lines, and with previous IRAC observations. Variations in the PAH ratios such as $6.2/11.3\mu$m were observed across the disk. However, they are strongly affected by the silicate feature at $10\mu$m.
We studied the variations the $6.2/7.7\mu$m and $11.3/7.7\mu$m PAH ratios, which are diagnostics for the size and the degree of ionization of PAHs.
We found no correlation between the $6.2/7.7\mu$m ratio with radiation hardness or with any other resolved
spatial structure in the central region. Since the variations are comparable to the uncertainties in the measurement the results are not (yet) significant
enough to support PAH size variations at parsec scales.
Similarly, the variations of the $11.3/7.7\mu$m ratio are mostly due to extinction and show little support for variations in PAH ionization throughout the region.

The $17\mu$m PAH complex is very prominent in the center of M82. We did not find any relative variations within the complex, which is in agreement with predictions. The variations of the $17/6.2\mu$m ratio are most likely due to extinction effects. Due to the uncertainties on the determination of extinction, we did not consider the remaining variations as a clear indicator of PAH size variation. 

We observed an enhancement of the EWs of the $11.3\mu$m and the other PAH features longwards of $10\mu$m outwards from the galactic plane. 
Several explanations exist for this, but we favor the variation of the UV radiation field, which excites differently PAHs and VSGs, given their 
different sizes.

The S(1) and S(2) rotational transition lines of H$_2$ have been detected in our spectra throughout the central region. H$_2$ and PAHs coincide at least on spatial 
scales of the resolution of the map, which is 35~pc.
We conclude that large scale shocks in the outflow regions have no measurable influence on the H$_2$ emission. However, we cannot 
distinguish between UV- and local
shock excitation as produced by supernovae or energetic outflows, on scales smaller than the resolution of our pixels (35 pc).

The Spitzer-IRS observations of the central region of M82 complements previous studies, not only in mid-infrared, but also in other wavelengths. 
Our results demonstrate the importance of spatially
resolved spectroscopy in starburst studies. They helped to constrain the age of the starburst and confirm results from other studies and also stressed the 
importance of a thorough study of extinction to investigate possible variations of PAH properties.
Further research of the starburst feedback and quenching processes will elucidate the sharp decrease in star formation in the last 5 Myr. 
It would be interesting to see if higher spatial resolution, as expected from the MIRI on board of JWST will discover a wider variation of the radiation field or signs of variation of PAH sizes an/or ionization. In addition, spatially more extended spectral studies of M82 are necessary to study the connection of the ionic line ratios with the outflows.

\acknowledgements
 
We would like to thank L. Snijders for making the Starburst99+Mappings models available. We would like to thank J. Lacy for providing the mid-infrared maps from IRTF ground-based observations. We also thank the anonymous referee whose many thoughtful and useful comments greatly improved this paper. This work is based on observations made with the \textit{Spitzer Space Telescope}, which is operated by the Jet Propulsion Laboratory, California Institute of Technology, under NASA contract 1407. 

\clearpage

\begin{landscape}
\begin{deluxetable}{ccccccccccc}
\tabletypesize{\scriptsize}
\tablewidth{0pt}
\tablecaption{Main Emission Line Strengths in units of $10^{-19}$W 
              cm$^{-2}$}
\tablehead{
\colhead{} & \colhead{[ArII]} & \colhead{[SIV]} & \colhead{[NeII]} & \colhead{[NeIII]} &
           \colhead{[SIII]} & \colhead{H$_{2}$ S(1)} &
           \colhead{H$_{2}$ S(2)} & \colhead{S(1)/S(2)} & \colhead{T (S(1)-S(2))} &
           \colhead{[NeIII]/[NeII]}\\ 
\colhead{Region} & \colhead{$6.99\mu$m} & \colhead{$10.5\mu$m} & \colhead{$12.8\mu$m} &
           \colhead{$15.6\mu$m} & \colhead{$18.7\mu$m} &
           \colhead{$17.0\mu$m} & \colhead{$12.3\mu$m} & \colhead{} & \colhead{(K)}
           & \colhead{}
}
\startdata
1 & 0.30$\pm$0.03 & 0.04$\pm$0.01 & 1.04$\pm$0.03 & 0.22$\pm$0.01 & 0.38$\pm$0.01 &
0.12$\pm$0.01 & 0.07$\pm$0.01 & 1.71$\pm$0.46 & 381$\pm$44 & 0.21$\pm$0.02 \\
2 & 7.82$\pm$0.23 & 0.31$\pm$0.01 & 25.3$\pm$0.9 & 3.42$\pm$0.09 & 8.5$\pm$0.19 &
0.23$\pm$0.04 & 0.25$\pm$0.03 & 0.92$\pm$0.30 & 586$\pm$117 & 0.13$\pm$0.01 \\
3 & 5.77$\pm$0.10 & 0.33$\pm$0.01 & 18.4$\pm$0.4 & 3.58$\pm$0.06 & 5.92$\pm$0.18 &
0.25$\pm$0.04 & 0.27$\pm$0.03 & 0.93$\pm$0.28 & 583$\pm$109 & 0.19$\pm$0.02\\
4 & 0.44$\pm$0.02 & 0.03$\pm$0.01 & 1.73$\pm$0.04 & 0.36$\pm$0.01 & 0.66$\pm$0.02 &
0.08$\pm$0.003 & 0.06$\pm$0.003 & 1.33$\pm$0.13 & 439$\pm$31 & 0.21$\pm$0.02 \\
Center & 17.1$\pm$0.47 & 0.72$\pm$0.08 & 55.4$\pm$0.97 & 8.53$\pm$0.15 & 18.6$\pm$0.37
& 0.51$\pm$0.07 & 0.67$\pm$0.07 & 0.76$\pm$0.21 & 700$\pm$129 & 0.15$\pm$0.01 \\
Total & 44.8$\pm$1.49 & 2.16$\pm$0.13 & 141.2$\pm$3.10 & 23.6$\pm$0.38 & 49.0$\pm$0.83
& 2.53$\pm$0.09 & 2.47$\pm$0.11 & 1.02$\pm$0.09 & 536$\pm$32 & 0.17$\pm$0.02 \\
\enddata
\label{tablines}
\end{deluxetable}
\clearpage
\end{landscape}

\begin{landscape}
\begin{deluxetable}{cccccccccc}
\tabletypesize{\scriptsize}
\tablecaption{Main PAH Feature Strengths as Measured with PAHFIT}
\tablecolumns{10}
\tablewidth{0pt}
\tablehead{
\colhead{} & \colhead{ 6.2$\mu$m } & \colhead{ 7.7$\mu$m } & 
\colhead{ 8.6$\mu$m } & \colhead{ 11.3$\mu$m } & \colhead{ 12.6$\mu$m } &
\colhead{14.2$\mu$m } & \colhead{17$\mu$m complex } & \colhead{F(17$\mu$m complex)/ } & \colhead{F(17$\mu$m complex)/ }\\
\colhead{} & \colhead{Flux\tablenotemark{a}} &
\colhead{Flux\tablenotemark{a}} & \colhead{Flux\tablenotemark{a}} &
\colhead{Flux\tablenotemark{a}} & \colhead{Flux\tablenotemark{a}} &
\colhead{Flux\tablenotemark{a}} & \colhead{Flux\tablenotemark{a}} & \colhead{F(6.2$\mu$m)} & \colhead{F(11.3$\mu$m)}\\
\colhead{Region} & \colhead{EW\tablenotemark{b}} &
\colhead{EW\tablenotemark{b}} & \colhead{EW\tablenotemark{b}} &
\colhead{EW\tablenotemark{b}} & \colhead{EW\tablenotemark{b}} &
\colhead{EW\tablenotemark{b}} & \colhead{EW\tablenotemark{b}}
}
\startdata
1 & 8.9$\pm$0.1 & 29.9$\pm$0.1 & 6.2$\pm$0.1 & 10.9$\pm$0.1 &
         6.0$\pm$0.1 & 0.62$\pm$0.01 & 4.2$\pm$0.03 & 0.47 & 0.39 \\ 
         & 0.308 & 0.254 & 0.437 & 2.27 & 2.06 & 2.13 & 1.97 \\
2 & 154.7$\pm$0.1 & 528.0$\pm$0.2 & 99.3$\pm$0.1 & 77.9$\pm$0.1 &
         64.5$\pm$0.1 & 6.1$\pm$0.1 & 34.6$\pm$0.1 & 0.22 &
         0.44\\ 
         & 1.10 & 0.866 & 0.848 & 0.660 & 0.734 & 0.717 &
         0.753 \\
3 & 136.3$\pm$0.1 & 478.2$\pm$0.2 & 83.6$\pm$0.1 & 73.7$\pm$0.1 &
         57.2$\pm$0.1 & 5.5$\pm$0.1 & 29.1$\pm$0.1 & 0.21 &
         0.39\\ & 0.827 & 0.884 & 0.865 & 0.784 & 0.877 & 0.850 & 
         0.720 \\
4 & 13.3$\pm$0.1 & 49.6$\pm$0.1 & 11.3$\pm$0.1 & 15.7$\pm$0.1 &
         7.7$\pm$0.1 & 0.67$\pm$0.10 & 7.6$\pm$0.1 & 0.57 &
         0.48\\ & 0.390 & 0.320 & 0.575 & 1.86 & 1.52 & 1.30 & 
         2.16 \\
Center & 375.0$\pm$0.2 & 1320.0$\pm$0.5 & 242.4.0$\pm$0.5 &
       202.5$\pm$0.6 & 156.3$\pm$0.3 & 14.3$\pm$0.1 & 99.0$\pm$0.2 & 0.26 &
       0.49\\ & 1.22 & 0.991 & 0.937 & 0.758 & 0.791 &
       0.750 & 0.907 \\
Total & 1158.0$\pm$1.0 & 4121.7$\pm$3.4 & 743.9$\pm$1.5 & 736.0$\pm$1.7
      & 535.8$\pm$0.7 & 52.2$\pm$0.2 & 327.8$\pm$0.5 & 0.28 &
      0.45\\ & 1.00 & 1.00 & 1.00 & 1.00 & 1.00 & 1.00 &
      1.00 \\ \enddata
\label{tabpahs}
\tablenotetext{a}{Flux in units of $10^{-19}$W cm$^{-2}$.  The quoted
errors are only the PAHFIT fit residuals. EWs are normalized to the EWs for the whole SH slit.}
\tablenotetext{b}{Equivalent width in units of $\mu$m}
\end{deluxetable}
\clearpage
\end{landscape}

\end{document}